\documentclass[12pt,draftcls,onecolumn]{IEEEtran}
\usepackage{amsfonts}
\usepackage{amsmath,amssymb}
\usepackage{graphicx}
\usepackage{caption2}
\usepackage{epsfig}
\usepackage[dvips]{color}

\newtheorem{assumption}{Assumption}
\newtheorem{lemma}{Lemma}

\newtheorem{theorem}{Theorem}
\newtheorem{remark}{Remark}

\newtheorem{algorithm}{Algorithm}

\begin{document}

\title{Distributed average tracking for multiple reference signals with general linear dynamics}
%
%
%

\author{
Yu~Zhao,
        Zhisheng~Duan, and Zhongkui~Li
\thanks{Y. Zhao, Z. Duan and Zhongkui Li are with the State Key Laboratory for Turbulence
and Complex Systems, Department of Mechanics and Engineering Science,
College of Engineering, Peking University, Beijing 100871, P. R. China
(e-mail: yuzhao5977@gmail.com; duanzs@pku.edu.cn; zhongkli@gmail.com).}
}


\maketitle

\begin{abstract} This technical note studies the distributed average tracking problem for multiple time-varying signals with general linear dynamics, whose reference inputs are nonzero and not available to any agent in the network. In distributed fashion, a pair of continuous algorithms with, respectively, static and adaptive coupling strengths are designed. Based on the boundary layer concept, the proposed continuous algorithm with static coupling strengths can asymptotically track the average of the multiple reference signals without chattering phenomenon. Furthermore, for the case of algorithms with adaptive coupling strengths, the average tracking errors are uniformly ultimately bounded and exponentially converge to a small adjustable bounded set. Finally, a simulation example is presented to show the validity of the theoretical results.
\end{abstract}

\section{Introduction}
In the past two decades, there has been lots of interest in distributed cooperative control for multi-agent systems due to its potential applications in formation flying, sensor networks, path planning and so forth. As a fundamental issue arising from distributed cooperative control for multi-agent systems, consensus has received compelling attention from various scientific communities \cite{OlfatiSaber1}, \cite{Ren:07}, \cite{Hong:08}, \cite{Cao:12}, \cite{Litac}, \cite{Zhang:11} and \cite{Ji}, ranging from mathematics to engineering, due to its less communication cost, greater efficiency, higher robustness, and so on.

In recent years, distributed average tracking, as a generalization of consensus and cooperative tracking problem, has received increasing attention and been applied in many different perspectives, such as distributed sensor networks \cite{Spanos2}, \cite{Bai2011} and distributed coordination \cite{Yang}, \cite{Sun}. The objective of distributed average tracking problem is to design a distributed algorithm for multi-agent systems to track the average of the multiple reference signals.
The motivation of this problem comes from coordinated tracking for multiple camera systems. Motivated by the pioneering works in \cite{Spanos1}, and \cite{Freeman} on distributed average tracking via linear algorithms, the applications of the related results on distributed sensor fusion \cite{Spanos2}, \cite{Bai2011}, and formation control \cite{Yang} were proceeded from the reality. In \cite{Bai}, distributed average tracking was investigated by considering the robustness to the initial errors in algorithms. The above-mentioned results are very interesting and important for scientific researchers to build up a general framework to investigate this topic. However, a common assumption in the above works is that the dynamics of the multiple reference signals are linear such as constant reference
signals \cite{Freeman} or reference signals with steady states \cite{Spanos1}. In practical applications, the reference signals may be produced by more general dynamics. Motivated by this observation, a class of nonlinear algorithms was designed in \cite{Nosrati:12} to track multiple reference signals with a bounded deviation. Then, based on the non-smooth control approaches, a couple of distributed algorithms were proposed in \cite{Chengfei:12} and \cite{Chengfei:13} for agents to track arbitrary time-varying reference signals with bounded derivatives and bounded accelerations, respectively. Further results in \cite{Chengfei:13ccc} studied the distributed average tracking problems for multiple bounded signals with linear dynamics.

Motivated by the above mentioned observations, this technical note is devoted to solving the distributed average tracking problem with continuous algorithms, for multiple time-varying signals with general linear dynamics, whose reference inputs are assumed to be nonzero and not available to any agent in the network. First of all, based on the relative states of the neighboring agents, a class of distributed continuous control algorithms is proposed and analyzed. Then, in light of adaptive control technique, a novel class of distributed algorithms with adaptive coupling strengths is designed.
Different from \cite{Cao:12}, \cite{Litac} and \cite{Chengfei:13ccc}, where the nonlinear signum function was applied to the whole neighborhood (node-based algorithm), while the proposed algorithms in this technical note are designed along the edge-based framework as in \cite{Chengfei:12} and \cite{Chengfei:13}.
Compared with the existing results, the contributions of this technical note are at least three-fold. First, two smooth control algorithms are proposed in this technical note, which are continuous approximations via the boundary layer concept and play a vital role to reduce chattering phenomenon in real applications. Second, the continuous distributed algorithms proposed in this technical note successfully solve the distributed average tracking problems for more general linear dynamics without the assumption of bounded signals as required in \cite{Chengfei:13ccc}. Third, from the viewpoint of consensus issues of heterogeneous uncertain  multi-agent systems, the value of consensus manifold can be obtained in this technical note. To
the best of our knowledge, it is the first time to give the expression of the consensus state.

\emph{Notations}: Let $R^n$ and $R^{n\times n}$ be the sets of real numbers and real matrices, respectively. $I_n$ represents the identity matrix of dimension $n$. Denote by $\mathbf{1}$ a column vector with
all entries equal to one. The matrix inequality $A> (\geq) B$ means that
$A-B$ is positive (semi-) definite. Denote by $A\otimes B$ the Kronecker product
of matrices $A$ and $B$. For a vector $x\in R^n$, let $\|x\|$
denote 2-norm of $x$. For a set $V$, $|V|$ represents the number of elements in $V$.

\section{Preliminaries}

\subsection{Graph Theory}
An undirected (simple) graph $\mathcal{G}$ is specified by a vertex set $\mathcal{V}$ and an edge set $\mathcal{E}$ whose elements characterize the incidence relation between distinct pairs of $\mathcal{V}$. The notation $i\sim j$ is used to denote that node $i$ is connected to node $j$, or equivalently, $(i, j)\in \mathcal{E}$. We make use of the $|\mathcal{V}|\times|\mathcal{E}|$ incidence matrix, $D(\mathcal{G})$, for a graph with an arbitrary orientation, i.e., a graph whose edges have a head (terminal node) and a tail (an initial node). The columns of $D(\mathcal{G})$ are then indexed by the edge set, and the $i$th row entry takes the value $1$ if it is the initial node of the corresponding edge, $-1$ if it is the terminal node, and zero otherwise. The diagonal matrix $\Delta(\mathcal{G})$ of the graph contains the degree of each vertex on its diagonal. The adjacency matrix, $A(\mathcal{G})$, is the $|\mathcal{V}|\times|\mathcal{V}|$ symmetric matrix with zero on the diagonal and one in the $(i,j)$th position if node $i$ is adjacent to node $j$. The graph Laplacian  \cite{GraphTheory} of $\mathcal{G}$, $L:= D(\mathcal{G})D(\mathcal{G})^T=\Delta(\mathcal{G})-A(\mathcal{G})$,
is a rank deficient positive semi-definite matrix.

An undirected path between node $i_1$ and node $i_s$ on undirected graph means a sequence of ordered undirected edges with the form $(i_k; i_{k+1}), k = 1, \cdots, s-1$.
A graph $\mathcal{G}$ is said to be connected if there exists a path between each pair of distinct nodes.

\begin{assumption}\label{ass}
Graph $\mathcal{G}$ is undirected and connected.
\end{assumption}

\begin{lemma} \label{lemma1}\cite{GraphTheory}
Under Assumption \ref{ass}, zero is a simple eigenvalue of $L$ with
$\mathbf{1}$ as an eigenvector and all the other eigenvalues are positive. Moreover, the
smallest nonzero eigenvalue $\lambda_2$ of L satisfies $\lambda_2=\min\limits_{ x\neq 0, \mathbf{1}^Tx=0 } \frac{x^TLx}{x^Tx}$.
\end{lemma}

\section{Distributed average tracking for multiple reference signals with general linear dynamics}

Suppose that there are $N$ time-varying reference signals, $r_i(t)\in R^n, i=1,2,\cdots, N$, which satisfy the following linear dynamics:
\begin{eqnarray}\label{L referencesignals}
\dot{r}_i(t)=Ar_i(t)+Bf_i(t),
\end{eqnarray}
where $A$ and $B$ are constant matrices with compatible dimensions, $r_i(t)$ is the state of the $i$th signal, and $f_i(t) $ represents the reference input of the $i$th signal.  Here, we assume that $f_i(t)$ is bounded and continuous,  i.e., $\|f_i(t)\|\leq f_0$, for $i=1,2,\cdots, N$, where $f_0$ is a positive constant. Suppose that there are $N$ agents with $x_i\in R^n$ being the state of the $i$th agent in a distributed algorithm. It is assumed that agent $i$ has access to $r_i(t)$, and agent $i$ can obtain the relative information from its neighbors denoted by $\mathcal{N}_i$, $i=1,2,\cdots, N$.

The main objective of this technical note is to design a distributed smooth algorithm for agents to track the average of multiple signals $r_i(t)$ described by general linear dynamics  (\ref{L referencesignals}) with bounded reference inputs $f_i(t),\;i=1,2,\cdots,N$.

Therefore, a distributed smooth algorithm is proposed as follows:
\begin{eqnarray}\label{L distributed control algorithm}
\dot{s}_i(t) &=&As_i(t)+c_1B\sum_{j\in \mathcal{N}_i}[K(x_i(t)-x_j(t))]+c_2B\sum_{j\in \mathcal{N}_i} h_i [K(x_i(t)-x_j(t))],\nonumber \\
{x}_i(t) &=& s_i(t)+r_i(t),\;s_i(0)=0,
\end{eqnarray}
where $s_i(t), \;i=1,2,\cdots,N$, are the internal states of the distributed filter (\ref{L distributed control algorithm}), $c_1$, $c_2$ and $K$ are coupling strengths and feedback gain matrix, respectively, to be determined, the nonlinear function $h_i(\cdot)$ are defined as follows: for $\omega\in R^n$,
\begin{eqnarray}\label{hi}
h_i(\omega)=\frac{\omega}{\|\omega\|+\varepsilon e^{-\varphi t} },
\end{eqnarray}
where $\varepsilon$ and $\varphi$ are positive constants.

Note that the nonlinear functions $h_i(\omega)$ in (\ref{hi}) are continuous,
which are actually continuous approximations, via the boundary layer concept \cite{28}, of the discontinuous
function
\begin{eqnarray*}
\widehat{h}_i(\omega)= \Bigg\{\begin{array}{cc}
                                                             \frac{\omega}{\|\omega\|} & \;\;\mathrm{if} \;\; \omega\neq 0, \\
                                                             0 & \;\;\mathrm{if} \;\;\omega= 0.\\
                                                           \end{array}
\end{eqnarray*}
The item $\varepsilon e^{-\varphi t}$ in (\ref{hi}) defines the sizes of the boundary layers. As $t\rightarrow \infty$, the continuous functions $h_i(\omega)$ approaches the discontinuous function $\widehat{h}_i(\omega)$.
It follows from (\ref{L referencesignals}) and (\ref{L distributed control algorithm}) that the closed-loop system is described by
\begin{eqnarray}\label{L closedloop}
\dot{x}_i(t)&=&Ax_i(t)+c_1B\sum_{j\in \mathcal{N}_i}[K(x_i(t)-x_j(t))]+c_2B\sum_{j\in \mathcal{N}_i} h_i [K(x_i(t)-x_j(t))]+Bf_i(t).
\end{eqnarray}

Before moving on, an important lemma is proposed.
\begin{lemma}\label{L lemma3}
Under Assumption \ref{ass}, the states $x_i(t)$ in (\ref{L distributed control algorithm}) will track the average of the multiple signals, i.e., $\|x_i(t)-\frac{1}{N}\sum_{k=1}^Nr_k(t)\|=0$, if the closed-loop system (\ref{L closedloop}) achieves consensus, i.e., $\lim_{t\rightarrow \infty}\|x_i-\frac{1}{N}\sum_{k=1}^N x_k\|=0$  for $i=1,2,\cdots, N$.
\end{lemma}
\textbf{Proof}: It follows from Assumption \ref{ass} that
\begin{eqnarray}\label{L1}
\sum_{i=1}^N\sum_{j\in \mathcal{N}_i}[K(x_i(t)-x_j(t))]=0, \;\;\sum_{i=1}^N\sum_{j\in \mathcal{N}_i} h_i [K(x_i(t)-x_j(t))]=0.
\end{eqnarray}
Let $S(t)=\sum_{i=1}^Nx_i(t)-\sum_{i=1}^Nr_i(t)$.  From (\ref{L referencesignals}), (\ref{L closedloop}) and (\ref{L1}), we have
\begin{eqnarray}\label{L2}
\dot{S}(t)=AS(t),
\end{eqnarray}
with $S(0)=0$.
By solving the differential equation (\ref{L2}) with initial condition above, we always have $S(t)=e^{At}S(0)=0$.
Thus, we obtain
\begin{eqnarray}\label{L3}
\sum_{i=1}^Nx_i(t)=\sum_{i=1}^Nr_i(t).
\end{eqnarray}
According to  Assumption \ref{ass}, if $x_i(t)$ in (\ref{L closedloop}) achieves consensus, i.e., $\lim_{t\rightarrow \infty}\|x_i(t)-\frac{1}{N}\sum_{k=1}^Nx_k(t)\|=0$ for $i=1,2,\cdots, N$, it follows from (\ref{L3}) that $\lim_{t\rightarrow \infty}\|x_i-\frac{1}{N}\sum_{k=1}^Nr_k(t)\|=0$, for $i=1,2,\cdots, N$. This completes the proof.
\begin{remark}
In the proof of Lemma \ref{L lemma3}, it requires that $ s_i(0)= 0$, which is a necessary condition to draw conclusions, if $A$ is not asymptotically stable. In the case that $A$ is asymptotically stable, without requiring the initial condition  $ s_i(0)= 0$, we can still reach the same conclusions as shown in Lemma \ref{L lemma3}, since the solution of (\ref{L2}) will converge to the origin for any initial condition.
\end{remark}

Let $x(t)=(x_1^T(t),x_2^T(t),\cdots,x_N^T(t))^T$, and $F(t)=(f_1^T(t),f_2^T(t),\cdots,f_N^T(t))^T$.
Define $\xi(t)=(M\otimes I)x(t)$, where $M=I_N-\frac{1}{N}\mathbf{1}\mathbf{1}^T$ and $\xi(t)=(\xi_1^T(t),\xi_2^T(t),\cdots,\xi_N^T(t))^T$. It is easy to see that $0$ is a simple eigenvalue
of $M$ with $\mathbf{1}$ as a corresponding right eigenvector and $1$ is the other eigenvalue with multiplicity $N-1$. Then, it follows that $\xi(t)= 0$ if and only if $x_1(t)=x_2(t)=\cdots=x_N(t)$. Therefore, the consensus problem of (\ref{L closedloop}) is solved if and only if $\xi(t)$ asymptotically converges to zero. Hereafter, we refer to $\xi(t)$ as the consensus error. By
noting that $LM = L$ and $MD(\mathcal{G})=D(\mathcal{G})$, it is not difficult to obtain from (\ref{L closedloop}) that the consensus error $\xi(t)$ satisfies
\begin{eqnarray}\label{LME closedloop}
\dot{\xi}(t)&=&(I{\otimes} A{+}c_1L{\otimes} BK)\xi(t){+}c_2(D(\mathcal{G}){\otimes} B)H[(D^T(\mathcal{G}){\otimes} K)\xi(t)]{+}(M{\otimes} B)F(t),
\end{eqnarray}
where
$
(D^T(\mathcal{G}){\otimes} B)H[(D^T(\mathcal{G}){\otimes} K)\xi(t)]=\left(
                                              \begin{array}{c}
                                                B\sum\limits_{j\in \mathcal{N}_1}h_1 [K(\xi_1(t)-\xi_j(t))] \\
                                                \vdots \\
                                                B\sum\limits_{j\in \mathcal{N}_N}h_N [K(\xi_N(t)-\xi_j(t))] \\
                                              \end{array}
                                      \right).
$

\begin{algorithm}\label{algorithm1}
For multiple reference signals in (\ref{L referencesignals}), the distributed average tracking algorithm (\ref{L distributed control algorithm}) can be constructed as follows
\begin{enumerate}
  \item Solve the algebraic Ricatti equation (ARE):
\begin{eqnarray}\label{LMI}
PA+A^TP-PBB^TP+Q=0,
\end{eqnarray}
with $Q>0$ to obtain a matrix $P>0$. Then, choose $K=-B^TP$.
  \item Select the first coupling strength $c_1\geq \frac{1}{2\lambda_{2}}$, where $\lambda_{2}$ is the smallest nonzero eigenvalue of the Laplacian $L$ of $\mathcal{G}$.
  \item Choose the second coupling strength $c_2\geq f_0(N-1)$, where $f_0$ is defined as in (\ref{L referencesignals}).
\end{enumerate}\end{algorithm}

\begin{theorem}\label{L theorem1f}
Under Assumption \ref{ass}, the states $x_i(t)$ in (\ref{L distributed control algorithm}) will track the average of multiple reference signals $r_i(t),\;i=1,2,\cdots, N$, described by general linear dynamics (\ref{L referencesignals}) with bounded reference inputs if the coupling strengths $c_1$, $c_2$ and the feedback gain $K$  are designed by Algorithm 1.
\end{theorem}
\textbf{Proof}:
Consider the Lyapunov function candidate
\begin{eqnarray}\label{LV 1}
V_1(t)= \xi^T(M\otimes P)\xi.
\end{eqnarray}
By the definition of $\xi(t)$, it is easy to see that $(\mathbf{1}^T\otimes I)\xi = 0$.
For a connected graph $\mathcal{G}$, it then follows from Lemma \ref{lemma1} that
\begin{eqnarray}\label{gLV 1}
V_1(t)\geq \lambda_{\min}(P)\|\xi\|^2.
\end{eqnarray}
The time derivative of $V_1$ along (\ref{LME closedloop}) can be obtained as follows
\begin{eqnarray}\label{dLV1}
\dot{ {V}}_1&=& \dot{\xi}^T(M\otimes P)\xi+ \xi^T(M\otimes P)\dot{\xi} \nonumber\\
&=& \xi^T(I\otimes A^T+c_1L\otimes K^TB^T)(M\otimes P)\xi +\xi^T(M\otimes P)(I\otimes A+c_1L\otimes BK)\xi\nonumber\\
&&+2 c_2\xi^T (D(\mathcal{G})\otimes PB)H[(D^T(\mathcal{G})\otimes K)\xi] +2\xi^T(M\otimes PB)F(t).
\end{eqnarray}
Substituting $K=-B^TP$ into (\ref{dLV1}), it follows from the fact $LM=ML=L$ that
\begin{eqnarray}\label{ddLV1}
\dot{ {V}}_1
&=& \xi^T(M\otimes (PA+A^TP)-2c_1L\otimes PBB^TP)\xi\nonumber\\
&&-2 c_2\xi^T (D(\mathcal{G})\otimes PB)H[(D^T(\mathcal{G})\otimes B^TP)\xi] +2\xi^T(M\otimes PB)F(t).
\end{eqnarray}
By using $\|F\|  \leq f_0$, we have
\begin{eqnarray}\label{lf5}
\xi^T(M\otimes PB)F(t)
&\leq& \|(M\otimes B^TP)\xi\| \|F(t)\| \nonumber\\
&\leq&
\frac{f_0}{N}\sum_{i=1}^N\sum_{j=1,j\neq i}^N\|B^TP(\xi_i-\xi_j)\|\nonumber\\
&\leq& \frac{f_0}{N}\sum_{i=1}^N \max_i\bigg\{\sum_{j=1,j\neq i}^N\|B^TP(\xi_i-\xi_j)\|\bigg\}\nonumber\\
&=& f_0 \max_i\bigg\{\sum_{j=1,j\neq i}^N\|B^TP(\xi_i-\xi_j)\|\bigg\}\nonumber\\
&\leq& \frac{f_0}{2}(N-1) \sum_{i=1}^N\sum_{j\in \mathcal{N}_i}
\|B^TP(\xi_i-\xi_j)\|.
\end{eqnarray}
Then, because of the facts that $\omega^Th_i(\omega)= \frac{\|\omega\|^2}{\|\omega\|+\varepsilon e^{-\varphi t}}$, we get
\begin{eqnarray}\label{lf6}
-2c_2\xi^T (D(\mathcal{G}){\otimes} PB)H[(D^T(\mathcal{G}){\otimes} B^TP)\xi]=-c_2 \sum_{i=1}^N\sum_{j\in \mathcal{N}_i} \frac{\|B^TP(\xi_i{-}\xi_j)\|^2}{\|B^TP(\xi_i{-}\xi_j)\|{+}\varepsilon e^{{-}\varphi t}}.
\end{eqnarray}
By combining with (\ref{lf5}) and (\ref{lf6}), it follows from (\ref{ddLV1}) that
\begin{eqnarray}\label{dddLV1}
\dot{ {V}}_1
&\leq&\xi^T(M{\otimes} (PA{+}A^TP){-}2c_1L{\otimes} PBB^TP)\xi+f_0(N-1) \sum_{i=1}^N\sum_{j\in \mathcal{N}_i} \|B^TP(\xi_i-\xi_j)\|\nonumber\\
&&-c_2 \sum_{i=1}^N\sum_{j\in \mathcal{N}_i}
\frac{\|B^TP(\xi_i-\xi_j)\|^2}{\|B^TP(\xi_i-\xi_j)\|+\varepsilon e^{-\varphi t}}.
\end{eqnarray}
Choose $c_2\geq f_0 (N-1)$. We have
\begin{eqnarray}\label{ddddLV1}
\dot{ {V}}_1
&\leq&\xi^T(M {\otimes} (PA{+}A^TP){-}2c_1L{\otimes} PBB^TP)\xi\nonumber\\
&&+c_2 \sum_{i=1}^N\sum_{j\in \mathcal{N}_i}
\Big(\|B^TP(\xi_i-\xi_j)\|-\frac{\|B^TP(\xi_i-\xi_j)\|^2}{\|B^TP(\xi_i-\xi_j)\|+\varepsilon e^{-\varphi t}}\Big)\nonumber\\
&\leq&\xi^T(M {\otimes} (PA{+}A^TP){-}2c_1L{\otimes} PBB^TP)\xi+c_2 \sum_{i=1}^N\sum_{j\in \mathcal{N}_i}
\varepsilon e^{-\varphi t}.
\end{eqnarray}
Since Assumption \ref{ass}, there exists a unitary matrix $U$ thus that $L=U^T\Lambda U$, where $\Lambda=\mathrm{diag}(\lambda_1,\lambda_2,$\\
$\cdots,\lambda_N)$. Without loss of generality, assume that $0=\lambda_1<\lambda_2\leq\cdots\leq\lambda_N$. Thereby, following from the fact that $M^2=M$, we obtain
\begin{eqnarray}\label{dddddLV1}
\xi^T(M {\otimes} (PA{+}A^TP){-}2c_1L{\otimes} PBB^TP)\xi
\leq\xi^T(M \otimes (PA+A^TP-2c_1\lambda_2 PBB^TP))\xi.
\end{eqnarray}
Select $c_1\geq \frac{1}{2\lambda_2}$. It follows from (\ref{LMI}) that $PA+A^TP-2c_1\lambda_2 PBB^TP\leq-Q$. Therefore, we have
\begin{eqnarray}\label{ddddddLV1}
\dot{ {V}}_1&<&- \gamma V_1+c_2 \sum_{i=1}^N\sum_{j\in
\mathcal{N}_i} \varepsilon e^{-\varphi t},
\end{eqnarray}
where $\gamma=\frac{\lambda_{\min}(Q)}{\lambda_{\max}(P)}$. Thus, we obtain that
\begin{eqnarray}\label{ddddddoV1}
0{\leq}{ {V}}_1(t) {\leq} e^{-\gamma t}V_1(0){+}c_2\sum_{i=1}^N\sum_{j\in \mathcal{N}_i}\int_0^t\varepsilon e^{-\gamma(t-\tau){-}\varphi \tau}d\tau.\end{eqnarray}
By noting that
\begin{eqnarray}\label{dddddddoV1}
\int_0^t\varepsilon e^{-\gamma(t-\tau)-\varphi \tau}d\tau = \Bigg\{\begin{array}{cc}
                                                             \varepsilon te^{-\gamma t} & \;\;\mathrm{if} \;\; \gamma=\varphi, \\
                                                             \frac{\varepsilon}{\gamma-\varphi}(e^{-\varphi t}-e^{-\gamma t}) & \;\;\mathrm{if} \;\;\gamma \neq \varphi,\\
                                                           \end{array}\nonumber
\end{eqnarray}
we have that $V_1(t)$ will converge to origin as $t\rightarrow \infty$, which means that the states of (\ref{L closedloop}) will achieve consensus. Then, according to Lemma \ref{L lemma3}, we have that the tracking errors $\xi_i, i=1,2,\cdots,N$ satisfy
\begin{eqnarray}\label{trackingerror}
\lim_{t\rightarrow \infty}\xi_i(t)&=&\lim_{t\rightarrow \infty}\bigg(x_i(t)-\frac{1}{N}\sum_{k=1}^{N}x_k(t)\bigg)\nonumber\\
&=&\lim_{t\rightarrow \infty}\bigg(x_i(t)-\frac{1}{N}\sum_{k=1}^{N}r_k(t)\bigg)\nonumber\\
&=&0,
\end{eqnarray}
Therefore, the distributed average tracking problem is solved. This completes the proof.

\begin{remark}
It is worth mentioning that a necessary and sufficient condition for
the existence of $P>0$ to the ARE (\ref{LMI}) is that $(A,B)$ is stabilizable.
Therefore, the stabilizability of (A, B) is also a sufficient condition for the existence of (\ref{L distributed control algorithm}). In addition, the feedback gains $K$, $c_1$ and $c_2$ in (\ref{L distributed control algorithm}) can be independently designed in Algorithm 1.
\end{remark}



\begin{remark}
Note that (\ref{L closedloop}) can also be seen as a  heterogeneous matching uncertain linear multi-agent systems as shown in  \cite{Litac}, \cite{28}:
\begin{eqnarray}\label{multiagent}
\dot{x}_i(t)=Ax_i(t)+B(u_i(t)+f_i(t)),
\end{eqnarray}
by designing an edge-based consensus protocol as follows,
\begin{eqnarray}\label{ui}
u_i&=&c_1\sum_{j\in \mathcal{N}_i}[K(x_i(t)-x_j(t))]+c_2\sum_{j\in \mathcal{N}_i} h_i [K(x_i(t)-x_j(t))].
\end{eqnarray}
From the proof of Theorem \ref{L theorem1f}, we get that multi-agent systems (\ref{multiagent}) with (\ref{ui}) will achieve consensus. According to Lemma \ref{L lemma3},  $x_i(t)\rightarrow \frac{1}{N}\sum_{i=1}^Nr_i(t)$, as $t\rightarrow \infty$. Notice that  $r_i(t)=e^{At}r_i(0)+\int_0^te^{A(t-\tau)}f_i(\tau)d\tau$ and $r_i(0)=x_i(0)$, which implies
\begin{eqnarray}\label{consensus maniford}
x_i{\rightarrow}\frac{1}{N}(\mathbf{1}^T\otimes e^{At})\left[\left(
                                                       \begin{array}{c}
                                                         x_1(0) \\
                                                         \vdots \\
                                                         x_N(0)\\
                                                       \end{array}
                                                     \right)
{+}\left(
\begin{array}{c}
\int_0^te^{-A\tau}f_1(\tau)d\tau \\
\vdots \\
\int_0^te^{-A\tau}f_N(\tau)d\tau \\
\end{array}
\right)\right],
\end{eqnarray}
as $t\rightarrow \infty$.
Therefore, we get the value of consensus manifold of multi-agent systems (\ref{multiagent}) with distributed protocol (\ref{ui}).
To the best of our knowledge, it is the first time to obtain the value of consensus manifold for the proposed heterogeneous matching uncertain multi-agent systems (\ref{multiagent}).
\end{remark}

\section{Distributed average tracking with distributed adaptive coupling strengths}
Note that in the last section, the first coupling strength $c_1$, designed as $c_1>\frac{1}{2\lambda_2 }$ relies on the communication topology. The second coupling strength $c_2$, designed as $c_2>f_0(N-1)$, requires $f_0$ and $N$. Generally, the smallest nonzero eigenvalue $\lambda_2$, the number $N$ of vertex set $\mathcal{V}$ and the supper bound $f_0$ of $f_i(t)$ are global information, which are difficult to be obtained for each agent when the scale of the network is very large. Therefore, to overcome these restrictions, a distributed average tracking algorithm with distributed adaptive coupling strengths is proposed as follows:
\begin{eqnarray}\label{A distributed control algorithm}
\dot{s}_i(t) &=&As_i(t)+B\sum_{j\in \mathcal{N}_i}\alpha_{ij}(t)[K(x_i(t)-x_j(t))]+B\sum_{j\in \mathcal{N}_i} \beta_{ij}(t)h_i [K(x_i(t)-x_j(t))],\nonumber \\
{x}_i(t) &=& s_i(t)+r_i(t),\;s_i(0)=0,
\end{eqnarray}
with distributed adaptive laws
\begin{eqnarray}\label{A distributed adaptive law}
\dot{\alpha}_{ij}(t)&=&\mu[-\vartheta \alpha_{ij}(t)+(x_i(t)-x_j(t))^T \Gamma (x_i(t)-x_j(t))], \nonumber\\
\dot{\beta}_{ij}(t) &=&
\nu\Bigg[-\chi \beta_{ij}(t)+\frac{\|K(x_i(t)-x_j(t))\|^2}{\|K(x_i(t)-x_j(t))\|+\varepsilon e^{-\varphi t}}\Bigg],
\end{eqnarray}
where $\alpha_{ij}(t)$ and $\beta_{ij}(t)$ are two adaptive coupling strengths satisfying ${\alpha}_{ij}(0)={\alpha}_{ji}(0)$ and ${\beta}_{ij}(0)={\beta}_{ji}(0)$, $\Gamma\in R^{n\times n}$ is a constant gain matrix, $\mu$, $\nu$, $\vartheta$ and $\chi$ are positive constants.

It follows from (\ref{L referencesignals}) and (\ref{A distributed control algorithm}) that the closed-loop system is described by
\begin{eqnarray}\label{A closedloop}
\dot{x}_i(t)=Ax_i(t){+}B\sum_{j\in \mathcal{N}_i}\alpha_{ij}(t)[K(x_i(t){-}x_j(t))]{+}B\sum_{j\in \mathcal{N}_i} \beta_{ij}(t)h_i[K(x_i(t){-}x_j(t))]{+}Bf_i(t),
\end{eqnarray}
where $\alpha_{ij}(t)$ and $\beta_{ij}(t)$  are given by (\ref{A distributed adaptive law}).

Similarly as in the last section, the following lemma is firstly given.
\begin{lemma}\label{A lemma3}
Under Assumption \ref{ass}, for algorithm (\ref{A distributed control algorithm}) with (\ref{A distributed adaptive law}), if $\lim_{t\rightarrow 0}\|x_i-\frac{1}{N}\sum_{k=1}^Nx_k\|=0,\;i=1,2,\cdots, N$, then $\lim_{t\rightarrow \infty}\|x_i-\frac{1}{N}\sum_{k=1}^Nr_k\|=0,\;i=1,2,\cdots, N$.
\end{lemma}
\textbf{Proof}: Since $\alpha_{ij}(0)=\alpha_{ji}(0)$ and $\beta_{ij}(0)=\beta_{ji}(0)$, it follows from (\ref{A distributed adaptive law}) that $\alpha_{ij}(t)=\alpha_{ji}(t)$ and $\beta_{ij}(t)=\beta_{ji}(t)$. From Assumption \ref{ass}, we have
$
\sum_{i=1}^N\sum_{j\in \mathcal{N}_i}\alpha_{ij}(t)[K(x_i(t)-x_j(t))]=0,$ and $
\sum_{i=1}^N\sum_{j\in \mathcal{N}_i}\beta_{ij}(t) h_i
[K(x_i(t)-x_j(t))]=0.$
Similar to the proof of Lemma \ref{L lemma3}, we can draw the conclusions in (\ref{L3}).
This completes the proof.

The following theorem shows the ultimate boundedness of the tracking error and the adaptive coupling strengths.
\begin{theorem}\label{A theorem1f}
Under the Assumption \ref{ass}, the tracking error $\xi$ defined in (\ref{trackingerror}) and the adaptive gains $\alpha_{ij}(t)$ and $\beta_{ij}(t)$ are uniformly ultimately
bounded, if the feedback gains $\Gamma$ and $K$ are designed as $\Gamma=PBB^TP$ and $K=-B^TP$, respectively, where $P>0$ is the unique solution to ARE (\ref{LMI}).
Furthermore, the following statements hold.
\begin{enumerate}
  \item For any $\vartheta$ and $\chi$, $\xi$, $\widetilde{\alpha}_{ij}$ and $\widetilde{\beta}_{ij}$ exponentially converge to the following bounded set
      \begin{eqnarray}\label{omiga1}
\Omega_1&\triangleq &\Bigg\{\xi, \widetilde{\alpha}_{ij}(t), \widetilde{\beta}_{ij}(t): V_2<\frac{1}{\delta}\sum_{i=1}^N\sum_{j\in \mathcal{N}_i}\Big(\vartheta\frac{\overline{\alpha}^2}{2}+\chi\frac{\overline{\beta}^2}{2}\Big)\Bigg\},
\end{eqnarray}
where $\delta\leq\min\{\gamma, \mu\vartheta, \nu\chi\}$,
\begin{eqnarray}\label{AV 2}
V_2 =\xi^T(M\otimes P)\xi+\sum_{i=1}^N\sum_{j\in \mathcal{N}_i}\bigg(\frac{\widetilde{\alpha}_{ij}(t)^2}{2\mu}{+}\frac{\widetilde{\beta}_{ij}(t)^2}{2\nu}\bigg),
\end{eqnarray}
$\widetilde{\alpha}_{ij}(t)= \alpha_{ij}(t){-}\overline{\alpha} $, $\widetilde{\beta}_{ij}(t)= \beta_{ij}(t){-}\overline{\beta}$, $\overline{\alpha}\geq \frac{1}{2\lambda_{2}}$ and $\overline{\beta}\geq f_0 (N-1)$.
  \item If select $\vartheta$ and $\chi$ small enough, such that $\varrho \triangleq\max\{\mu\vartheta, \nu\chi\}<\gamma$, the tracking errors $\xi$ will exponentially converge to the bounded set $\Omega_2$ given as follows:
      \begin{eqnarray}\label{omiga}
\Omega_2\triangleq \bigg\{\xi: \|\xi\|\leq\bigg(\sum_{i=1}^N\sum_{j\in \mathcal{N}_i}\frac{\vartheta \overline{\alpha}^2 +\chi \overline{\beta}^2}{2\lambda_{\min}(P)(\gamma-\varrho)}\bigg)^{\frac{1}{2}}\bigg\},
\end{eqnarray}
where $\gamma$ is defined in (\ref{ddddddLV1}).
\end{enumerate}
\end{theorem}
\textbf{Proof}:
Consider the Lyapunov function candidate $V_2$ in (\ref{AV 2}).
As shown in the proof of Theorem \ref{L theorem1f}, the time derivative of $V_2$ along (\ref{A distributed adaptive law}) and (\ref{A closedloop}) satisfies
\begin{eqnarray}\label{dddLV2}
\dot{{ {V}}}_2
&\leq&\xi^T[M\otimes (PA+A^TP)]\xi -\sum_{i=1}^N\sum_{j\in \mathcal{N}_i}\alpha_{ij}(t)(\xi_i-\xi_j)^T PBB^TP(\xi_i-\xi_j)\nonumber\\
&&+ \sum_{i=1}^N\sum_{j\in \mathcal{N}_i}\bigg( f_0(N-1)\|B^TP(\xi_i-\xi_j)\| - \beta_{ij}(t)\frac{\|B^TP(x_i(t){-}x_j(t))\|^2}{\|B^TP(x_i(t){-}x_j(t))\|{+}\varepsilon e^{-\varphi t}}\bigg)\nonumber\\
&&+\frac{1}{\mu}\sum_{i=1}^N\sum_{j\in \mathcal{N}_i} \widetilde{\alpha}_{ij}(t)\dot{\alpha}_{ij}(t) + \frac{1}{\nu}\sum_{i=1}^N\sum_{j\in \mathcal{N}_i}\widetilde{\beta}_{ij}(t)\dot{\beta}_{ij}(t).
\end{eqnarray}
By using $\Gamma=PBB^TP$, it follows from (\ref{A distributed adaptive law}) that
\begin{eqnarray}\label{A2}
&&-\sum_{i=1}^N\sum_{j\in \mathcal{N}_i}\alpha_{ij}(t)(\xi_i-\xi_j)^T PBB^TP(\xi_i-\xi_j) +\frac{1}{\mu}\sum_{i=1}^N\sum_{j\in \mathcal{N}_i}  \widetilde{\alpha}_{ij}(t) \dot{\alpha}_{ij}(t)\nonumber\\
&\leq& -2\overline{\alpha} \xi^T(L\otimes PBB^TP)\xi +\vartheta\sum_{i=1}^N\sum_{j\in \mathcal{N}_i} \Big(-\frac{\widetilde{\alpha}_{ij}(t)^2}{2}+\frac{\overline{\alpha}^2}{2}\Big),
\end{eqnarray}
and
\begin{eqnarray}\label{A3}
&&- \sum_{i=1}^N\sum_{j\in \mathcal{N}_i}
\beta_{ij}(t)\frac{\|B^TP(x_i(t)-x_j(t))\|^2}{\|B^TP(x_i(t)-x_j(t))\|+\varepsilon e^{-\varphi t}} +\frac{1}{\nu}\sum_{i=1}^N\sum_{j\in \mathcal{N}_i} \widetilde{\beta}_{ij}(t) \dot{\beta}_{ij}(t)\nonumber\\
&\leq& -\overline{\beta}\sum_{i=1}^N\sum_{j\in
\mathcal{N}_i}\frac{\|B^TP(x_i(t)-x_j(t))\|^2}{\|B^TP(x_i(t)-x_j(t))\|+\varepsilon e^{-\varphi t}} +\chi\sum_{i=1}^N\sum_{j\in \mathcal{N}_i}\Big(-\frac{\widetilde{\beta}_{ij}(t)^2}{2}+\frac{\overline{\beta}^2}{2}\Big).
\end{eqnarray}
Substituting (\ref{A2}) and (\ref{A3}) into (\ref{dddLV2}), we have
\begin{eqnarray}\label{dddAV2}
\dot{{ {V}}}_2
&\leq&\xi^T(M {\otimes} (PA+A^TP){-}2\overline{\alpha}L{\otimes} PBB^TP)\xi +f_0(N-1) \sum_{i=1}^N\sum_{j\in \mathcal{N}_i} \|B^TP(\xi_i-\xi_j)\|\nonumber\\
&&-\overline{\beta} \sum_{i=1}^N\sum_{j\in \mathcal{N}_i}
\frac{\|B^TP(x_i(t)-x_j(t))\|^2}{\|B^TP(x_i(t)-x_j(t))\|+\varepsilon e^{-\varphi t}} +\sum_{i=1}^N\sum_{j\in \mathcal{N}_i}\Bigg[ \vartheta\Big(-\frac{ \widetilde{\alpha}_{ij}(t) ^2}{2}+\frac{\overline{\alpha}^2}{2}\Big)\nonumber\\
&&+\chi\Big(-\frac{\widetilde{\beta}_{ij}(t)^2}{2}+\frac{\overline{\beta}^2}{2}\Big)\Bigg].
\end{eqnarray}
As shown in the proof of Theorem \ref{L theorem1f}, by choosing $\overline{\alpha}$ and $\overline{\beta}$ sufficiently large such that  $\overline{\alpha}\geq \frac{1}{2\lambda_{2}}$ and $\overline{\beta}\geq f_0 (N-1)$, we have
\begin{eqnarray}\label{AAA v}
\dot{{{V}}}_2 &\leq& -\xi^T(M {\otimes} (PA+A^TP{-} PBB^TP))\xi +\overline{\beta} \sum_{i=1}^N\sum_{j\in \mathcal{N}_i}\varepsilon e^{-\varphi t}+\sum_{i=1}^N\sum_{j\in \mathcal{N}_i}\Big(\vartheta\frac{\overline{\alpha}^2}{2}+\chi\frac{\overline{\beta}^2}{2}\Big)\nonumber\\
&&-\sum_{i=1}^N\sum_{j\in \mathcal{N}_i}\Bigg( \vartheta\frac{ \widetilde{\alpha}_{ij}(t)^2}{2}+\chi\frac{\widetilde{\beta}_{ij}(t)^2}{2}\Bigg).
\end{eqnarray}
Since $\delta\leq\min\{\gamma, \mu\vartheta, \nu\chi\}$, we obtain that
\begin{eqnarray}\label{AAA vaaa}
\dot{{{V}}}_2&\leq& {-}\delta V_2{+}\sum_{i=1}^N\sum_{j\in \mathcal{N}_i}\frac{(\delta{-}\mu\vartheta) \widetilde{\alpha}_{ij}(t)^2}{2\mu}{+}\frac{(\delta{-}\nu\chi)\widetilde{\beta}_{ij}(t)^2}{2\nu}
\nonumber\\
&&+\overline{\beta}
 \sum_{i=1}^N\sum_{j\in \mathcal{N}_i}\varepsilon e^{-\varphi t}+\sum_{i=1}^N\sum_{j\in \mathcal{N}_i}\Big(\vartheta\frac{\overline{\alpha}^2}{2}+\chi\frac{\overline{\beta}^2}{2}\Big)\nonumber\\
&\leq&-\delta V_2+\overline{\beta}
 \sum_{i=1}^N\sum_{j\in \mathcal{N}_i}\varepsilon e^{-\varphi t} +\sum_{i=1}^N\sum_{j\in \mathcal{N}_i}\Big(\vartheta\frac{\overline{\alpha}^2}{2}+\chi\frac{\overline{\beta}^2}{2}\Big).
\end{eqnarray}
In light of the well-known Comparison lemma in \cite{kalia}, we can obtain from (\ref{AAA vaaa}) that
\begin{eqnarray}\label{ddddddoV2}
{ {V}}_2(t) &{\leq}& e^{-\delta t}\Big[V_2(0)+\frac{1}{\delta}\sum_{i=1}^N\sum_{j\in \mathcal{N}_i}\Big(\vartheta\frac{\overline{\alpha}^2}{2}+\chi\frac{\overline{\beta}^2}{2}\Big)\Big] {+}\overline{\beta}\sum_{i=1}^N\sum_{j\in \mathcal{N}_i}\int_0^t\varepsilon e^{-\delta(t-\tau){-}\varphi \tau}d\tau \nonumber\\
&&+\frac{1}{\delta}\sum_{i=1}^N\sum_{j\in \mathcal{N}_i}\Big(\vartheta\frac{\overline{\alpha}^2}{2}+\chi\frac{\overline{\beta}^2}{2}\Big).\end{eqnarray}
Therefore, $V_2(t)$ exponentially converges to the bounded set $\Omega_1$ as given in (\ref{omiga1}).
It implies that $\xi(t)$, $\alpha_{ij}(t)$ and $\beta_{ij}(t)$ are uniformly ultimately bounded.

Next, if $\varrho \triangleq\max\{\mu\vartheta, \nu\chi\}<\gamma$, we can obtain a smaller set for $\xi$
by rewriting (\ref{AAA v}) into
\begin{eqnarray}\label{AAA v2}
\dot{{{V}}}_2 &\leq& -\varrho V_2-\lambda_{\min}(P)(\gamma-\varrho)\|\xi\|^2
 +
 \sum_{i=1}^N\sum_{j\in \mathcal{N}_i}\bigg[\overline{\beta}\varepsilon e^{-\varphi t}{+}\Big(\vartheta\frac{\overline{\alpha}^2}{2}{+}\chi\frac{\overline{\beta}^2}{2}\Big)\bigg].
\end{eqnarray}
Obviously, it follows from (\ref{AAA v2}) that $\dot{V}_2(t)\leq -\varrho V_2(t)+\sum_{i=1}^N\sum_{j\in \mathcal{N}_i}\overline{\beta}\varepsilon e^{-\varphi t}$, if
$
\|\xi\|^2{>}\frac{1}{2\lambda_{\min}(P)(\gamma-\varrho)}\\
\sum_{i=1}^N\sum_{j\in \mathcal{N}_i}\Big(\vartheta \overline{\alpha}^2 +\chi \overline{\beta}^2 \Big).
$
Then, in light of $V_2(t)\geq\lambda_{\min}(P)\|\xi\|^2$, we can get that if $\varrho<\gamma$ then $\xi$ exponentially converges to the bounded set $\Omega_2$ in (\ref{omiga}).
Therefore, we obtain from Lemma \ref{A lemma3} that distributed average tracking errors $\xi_i=x_i-\frac{1}{N}\sum_{k=1}^Nr_{k},\;i=1,2,\cdots,N$, converge to the bounded set $\Omega_2$ as $t\rightarrow \infty$. This completes the proof.

\begin{remark}
The adaptive scheme of algorithm (\ref{A distributed adaptive law}) for updating the coupling gains is partly borrowed from the adaptive strategies in \cite{Litac}, \cite{adaptive2} and \cite{adaptive3}. In Algorithm \ref{algorithm1}, it requires the smallest nonzero eigenvalue $\lambda_2$ of $L$, the upper bound $f_0$ of $f_i(t)$ and the number $N$ of nodes in the network. Note that $\lambda_2$, $f_0$ and $N$ are global information for each agent in the network and might not be obtained in real applications. By using adaptive strategies (\ref{A distributed control algorithm}) with (\ref{A distributed adaptive law}) in Theorem \ref{A theorem1f}, the limitation of all these global information can be removed.
\end{remark}
\begin{remark}
Note that the related works in \cite{Chengfei:12} and \cite{Chengfei:13} firstly studied the distributed average tracking problem for first-order and second-order integrators, respectively, by using non-smooth algorithms, which inevitably produces the chattering phenomenon. Compared with the results in \cite{Chengfei:12} and \cite{Chengfei:13}, the contribution of this technical note is at least three-fold. First, two smooth algorithms are proposed, which are continuous approximations via the boundary layer concept and play a vital role to reduce chattering phenomenon in real applications.
Second, it is assumed in \cite{Chengfei:13ccc} that the states of the reference signals $r_i(t)$ are bounded, which is not satisfied for some general signals such as ones with higher-order integrator-type dynamics. Generally, the continuous algorithms (\ref{L distributed control algorithm}) and (\ref{A distributed control algorithm}) proposed in this technical note successfully solve the distributed average tracking problems for more general linear dynamics without
the assumption as required in \cite{Chengfei:13ccc}. Third, from the viewpoint of consensus issues for heterogeneous uncertain  multi-agent systems, the value of consensus manifold can be obtained in this technical note, which is mainly attributed to the edge-based design patterns.
\end{remark}

\section{Simulations}
In this section, we will give an example to verify Theorem \ref{A theorem1f}.  The dynamics of multiple reference signals are given by (\ref{L referencesignals}) with
$
r_i=\left(
  \begin{array}{c}
     {r}_{1i} \\
     {r}_{2i} \\
  \end{array}
\right),\;\;A=\left(
          \begin{array}{cc}
            0 & 1 \\
            -1 & -2 \\
          \end{array}
        \right),\;\;B=\left(
  \begin{array}{c}
    0 \\
    1 \\
  \end{array}
\right),$
and $f_i(t)=\frac{i+1}{2} \sin(t)$, where $i=1,2,\cdots, 6$.
The communication topology is shown in Fig. 1.
Solving the ARE (\ref{LMI}) with $Q=I$ gives the
gain matrices $K$ and $\Gamma$ as
$
K{=}\left(
  \begin{array}{cc}
    -1.5728 \; &  -4.3293 \\
  \end{array}
\right), \\ \Gamma{=}
\left(
  \begin{array}{cc}
    2.4738 \; &  6.8092\\
    6.8092 \; & 18.7428\\
  \end{array}
\right).$
The tracking error trajectories $\xi_i=  x_i-\frac{1}{6}\sum_{k=1}^6r_k$, $i=1,2,\cdots,6$, of the six agents with respect to the average of the multiple reference signals with $\mu=10,\;\nu=10,\;\vartheta=0.01,\;\chi=0.01,\;\varepsilon=5,\;\varphi=0.5,\; K$ and $\Gamma$ given above are depicted in Fig. 2, which shows that the states indeed achieve average tracking. The adaptive coupling gains $\alpha_{ij}(t)$ and $\beta_{ij}(t)$ are also drawn in Fig. 3.
\begin{figure}
  \center{
  \includegraphics[width=3cm]{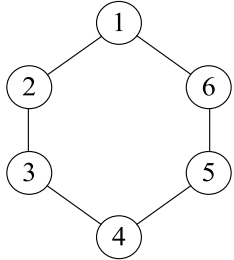}\\
  \caption{The communication topology.} }
\end{figure}
%
\begin{figure}
  \begin{minipage}[t]{0.5\linewidth}
\centering
  \includegraphics[width=6cm]{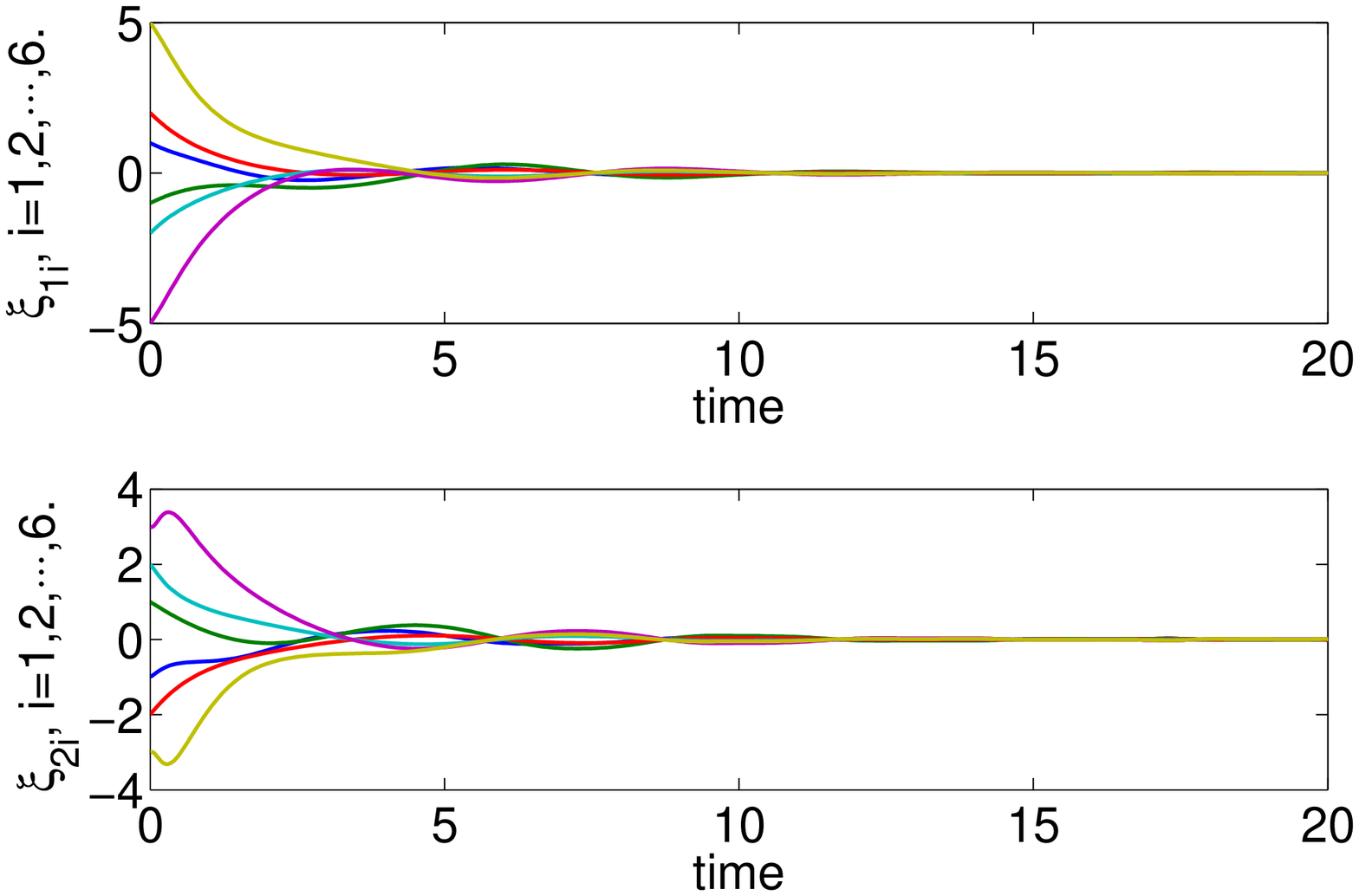}
  \caption{Tracking error trajectories $\xi_i$
   of the six agents in the network.}
  \end{minipage}
\begin{minipage}[t]{0.5\linewidth}
\centering
  \includegraphics[width=6cm]{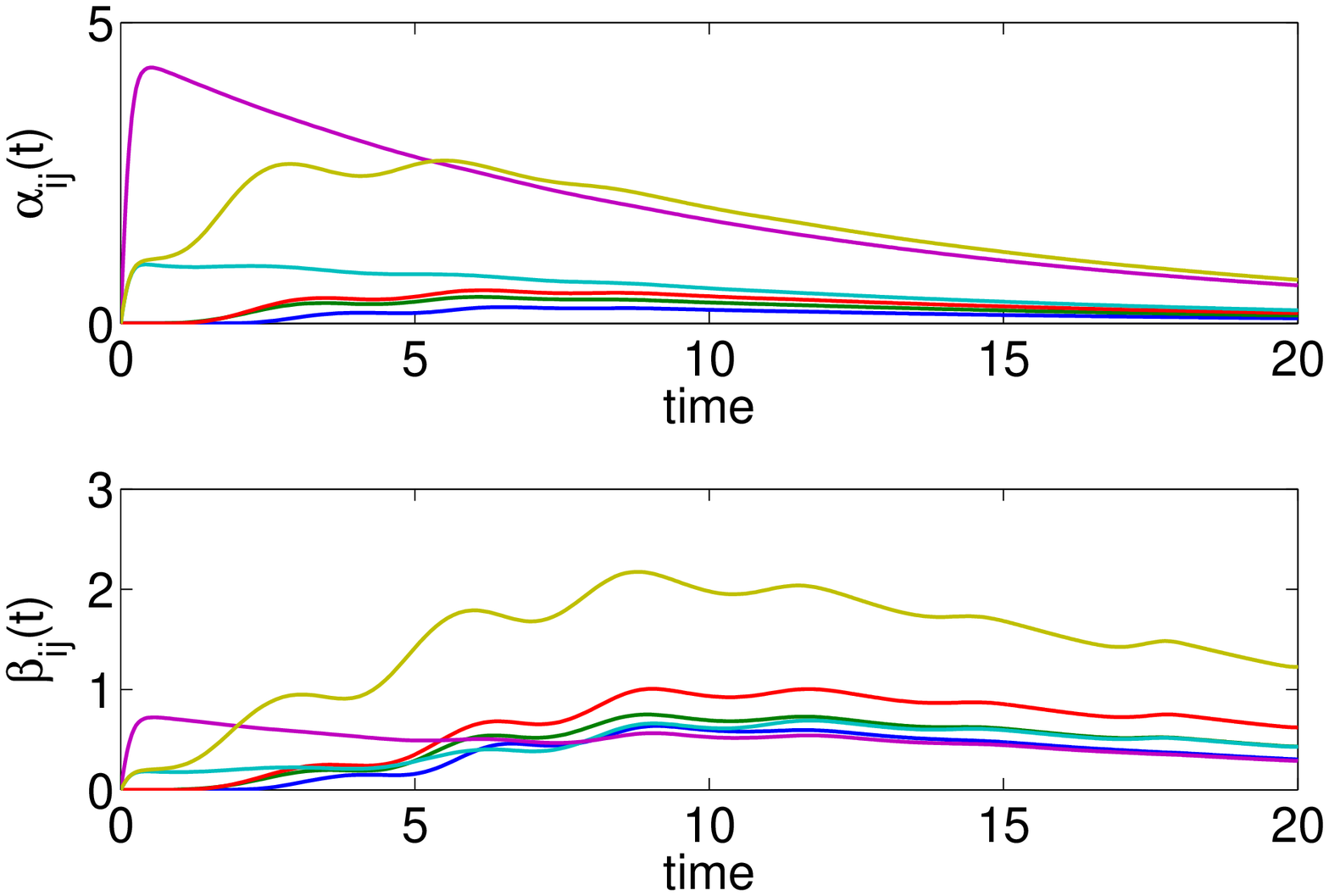}
  \caption{Adaptive coupling strengths $\alpha_{ij}(t)$ and $\beta_{ij}(t)$ in (\ref{A distributed adaptive law}).}
   \end{minipage}
\end{figure}
\section{Conclusions}
In this technical note, we have studied the distributed average tracking problem of multiple time-varying signals with general linear dynamics, whose reference inputs are nonzero, bounded and not available to any agents in networks. In distributed fashion, a pair of continuous algorithms with static and adaptive coupling strengths have been developed in light of the boundary layer concept. Besides, sufficient conditions for the existence of distributed algorithms are given if each agent is stabilizable. The future topic will be focused on distributed average tracking problem for the case with only relative output information of neighboring agents.
\linespread{1.4}


\end{document}